\documentclass[11pt,a4paper]{article}
\pdfoutput=1

\usepackage{lmodern} 
\usepackage[utf8]{inputenc} 
\usepackage[T1]{fontenc} 
\usepackage{microtype} 
\usepackage[english]{babel}

\usepackage{color,graphicx,subfigure}

\usepackage{amsmath,dsfont,url,amssymb}
\usepackage{slashed,cancel}
\usepackage[usenames,dvipsnames]{xcolor}
\usepackage[colorlinks=true, linkcolor=black, citecolor=Forest Green]{hyperref}
\usepackage{mdframed}
\usepackage{overpic}

\def \d {\partial}

\DeclareMathOperator{\Tr}{Tr}

\renewcommand{\Im}[0]{\operatorname{Im}}
\renewcommand{\Re}[0]{\operatorname{Re}}

\definecolor{inkred}{RGB}{210,29,0}
\definecolor{inkblue}{RGB}{0,112,196}

\numberwithin{equation}{section}

\begin{document}
\frenchspacing
\begin{center}

{ \Large {\bf
Theory of diffusive fluctuations
}}

\vspace{1cm}

Xinyi Chen-Lin,\, Luca V. Delacr\'etaz\, and\, Sean A. Hartnoll

\vspace{1cm}

{\small
{\it Department of Physics, Stanford University, \\
Stanford, CA 94305-4060, USA }}

\vspace{1.6cm}

\end{center}

\begin{abstract}

The recently developed effective field theory of fluctuations around thermal equilibrium is used to compute late-time correlation functions of conserved densities. Specializing to systems with a single conservation law, we find that the diffusive pole is shifted in the presence of non-linear hydrodynamic self-interactions, and that the density-density Green's function acquires a branch point half way to the diffusive pole, at frequency $\omega= -\frac{i}{2}Dk^2$. We discuss the relevance of diffusive fluctuations for strongly correlated transport in condensed matter and cold atomic systems.
\end{abstract}

\thispagestyle{empty}
\pagenumbering{arabic}

\pagebreak
\setcounter{page}{1}

\noindent
{\it La chaleur p\'en\`etre, comme la gravit\'e, toutes les substances de l'univers, ses rayons occupent toutes les parties de l'espace. \rm \hfill (Fourier, 1822)}

\section{Introduction}

Diffusion was invented by Fourier to describe the dynamics of heat \cite{fourier1822theorie}. Heat or energy transport is ubiquitous and of relevance to essentially any physical system at nonzero temperature. In modern parlance, diffusion is understood more generally as the universal late-time hydrodynamic description of systems governed by a single conservation law, barring spontaneously broken symmetries. The diffusion constant may itself depend on the background value of the quantity that is being transported. Classically, this results in the diffusion equation being augmented to a non-linear differential equation and can lead to rich phenomena (see e.g.~\cite{crank1979mathematics}), in analogy with turbulence in the Navier-Stokes equation. At the quantum or statistical physics level, the consequence of these non-linearities is that the effective theory of hydrodynamic fluctuations is interacting.

The traditional approach to address this class of problems is to couple the degrees of freedom of interest to stochastic noise fields and solve perturbatively a non-linear Langevin equation \cite{PhysRevA.8.423,POMEAU197563}. This approach is also familiar in the context of the KPZ equation \cite{PhysRevLett.56.889}. These methods have revealed striking effects in hydrodynamics such as long-time tails \cite{DESCHEPPER19741,POMEAU197563,Kovtun:2003vj,PhysRevB.73.035113} and potentially large renormalizations of transport parameters \cite{PhysRevA.16.732,Kovtun:2011np,Kovtun:2014nsa}. However, where standard classical hydrodynamic stood on firm symmetry principles \cite{chaikin1995principles}, the physical principles governing stochastic hydrodynamics -- in particular how `noise' fields interact with conserved densities -- were less transparent. This was remedied recently by a decade-long effort culminating in a first principles construction of the general effective field theory of hydrodynamic fluctuations about a thermal equilibrium state \cite{Dubovsky:2011sj,Grozdanov:2013dba,Haehl:2015pja,Crossley:2015evo,Jensen:2017kzi}, see \cite{Glorioso:2018wxw} for a recent review.

Another motivation for a systematic study of hydrodynamic fluctuations is thermalization. The local thermalization (or equilibration) time $\tau_{\rm th}$ is loosely defined as the time it takes a system to reach local thermodynamic equilibrium. At times $t>\tau_{\rm th}$, hydrodynamics governs the slower relaxation to global thermodynamic equilibrium.
It is tempting to identify the thermalization time with the exponential decay of non-hydrodynamic correlators $\langle \mathcal{O}(t) \mathcal{O}\rangle\sim e^{-t/\tau_{\rm th}}$. Such correlators are however sensitive in general to hydrodynamic long-time tails and therefore strictly do not decay exponentially \cite{POMEAU197563,Kovtun:2003vj}. A better understanding of long-time tails may therefore help provide a sharp definition of $\tau_{\rm th}$. See e.g.~\cite{Han:2018hlj,Lucas:2018wsc} for recent alternative approaches to $\tau_{\rm th}$.

In the following we use the general formalism of Ref.~\cite{Crossley:2015evo} to uncover the universal structure of late-time  response functions for interacting systems with a single continuous symmetry, focusing on time translation invariance (and therefore heat transport) for concreteness. We find that the thermal dc conductivity and diffusion constant both receive independent non-vanishing radiative corrections, even in the case of a single conserved density, and that the correction is not sign definite. Both of these statements are different to the results obtained from a traditional approach \cite{Kovtun:2014nsa}, for reasons we shall explain. Moreover, we compute the one-loop retarded Green's function $G^R_{\varepsilon \varepsilon}(\omega,k)$ at finite frequency and wavevector, revealing its analytic structure. We conclude by discussing experimental signatures of hydrodynamic fluctuations with applications to insulators, bad metals and cold atoms.

\section{Formalism}

Our objective is to understand the structure of energy density correlation functions in non-integrable quantum systems at nonzero temperature
\begin{equation}\label{eq_corr}
\langle \varepsilon (t, x) \varepsilon (t', x') \cdots\rangle_\beta
	\equiv \Tr \Bigl(\rho_{\beta}\,  \varepsilon (t,x) \varepsilon(t',x')\cdots \Bigr)\, ,
\end{equation}
where the thermal density matrix $\rho_\beta = e^{-\beta H}/\Tr e^{-\beta H}$. Here we will be interested in the case where energy is the only conserved quantity. The systematic study of a single diffusive charge was initiated in Ref.~\cite{Crossley:2015evo}. In that formalism, furthermore, the contribution of ghosts (or lack thereof) has been well understood \cite{Gao:2018bxz}. A self-contained review of this special case of the formalism is given in appendix \ref{app_Hong}. The output of this method is an effective field theory that provides a perturbative expansion for computing the correlators \eqref{eq_corr}:
\begin{equation}
Z = \int D \varepsilon D \varphi_a \, e^{i \int  \mathcal L[\varepsilon,\, \varphi_a]} \,.
\end{equation}
Here $\varepsilon$ is the energy density and $\varphi_a$ is an auxiliary field (the $a$ subscript is not an index). The most general Lagrangian to cubic order in fields was constructed in Ref.~\cite{Crossley:2015evo}. In appendix \ref{app_Hong} we extend their construction to quartic interactions, which will play a role below. The resulting Lagrangian to leading order in derivatives that is at most quartic in fields is given by
\begin{equation}\label{eq_action_main}
\mathcal L
	= iT^2\kappa (\nabla \varphi_a)^2 - \varphi_a \left(\dot \varepsilon -
	 D \nabla^2 \varepsilon\right) 
	+ \nabla^2\varphi_a\left[\frac12 \lambda\varepsilon^2 + \frac13\lambda'\varepsilon^3\right] +
	icT^2(\nabla\varphi_a)^2 \left[\widetilde\lambda\varepsilon + \widetilde\lambda'\varepsilon^2\right]
	+ \cdots \, ,
\end{equation}
where $T$ is the temperature, $D$ the diffusivity, $c$ the specific heat and $\kappa = c D$ the thermal conductivity. These are all `bare' values that will be renormalized by the interactions in (\ref{eq_action_main}). The couplings $\lambda,\lambda',\widetilde \lambda, \widetilde \lambda'$ themselves can be written as linear combinations of the following derivatives of the transport parameters
\begin{equation}\label{eq_couplings_raw}
T\d_T \kappa\, , \qquad\quad
T^2\d_T^2 \kappa\, , \qquad\quad
T\d_T D\, , \qquad\quad
T^2\d_T^2 D\, .
\end{equation}
Their explicit expressions are given in \eqref{eq_couplings}.

The traditional stochastic approach to hydrodynamic fluctuations with Gaussian noise \cite{PhysRevA.8.423,POMEAU197563,PhysRevLett.56.889,Kovtun:2003vj} can be recovered from the general effective action \eqref{eq_action_main} when the interactions that are quadratic in auxiliary fields (i.e. the $\widetilde\lambda$ and $\widetilde \lambda'$ terms) are absent, by performing a Legendre transform and introducing the noise field $\xi = \d \mathcal L /\d \varphi_a$ \cite{Crossley:2015evo}. However, when $\widetilde \lambda$ or $\widetilde \lambda'$ is non-vanishing, the resulting theory will contain  interactions of the form $\xi^2 \varepsilon$ or $\xi^2 \varepsilon^2$. The noise correlations will therefore not be strictly Gaussian because they now depend on energy fluctuations.%
	\footnote{That these interactions should arise is already clear from the stochatistic approach, where the fluctuation dissipation theorem imposes $\langle \xi(x)\xi\rangle = -2T^2\kappa(\varepsilon) \nabla^2 \delta(x)$. These interactions only vanish if $\kappa(\varepsilon) = \rm const$, which is also apparent in \eqref{eq_couplings}.} 

In the remainder we will show precisely how
the interactions in (\ref{eq_action_main}) lead to non-analyticities in response functions and renormalize the transport parameters themselves \cite{DESCHEPPER19741,PhysRevB.73.035113,PhysRevA.16.732,Kovtun:2011np,Kovtun:2012rj}. Concretely, we are interested in the one-loop correction to the retarded Green's function, which is simply diffusive in the absence of interactions
\begin{equation}\label{eq:bare}
G_{\varepsilon \varepsilon}^{R,0}(\omega,k)
	= \frac{i\kappa T k^2}{\omega + iDk^2}\, .
\end{equation}
The diagrams contributing at one loop are shown in Fig. \ref{fig_loops}, and computed in appendix \ref{app_loops}. These loops are all UV divergent and should be truncated at the hydrodynamic cutoff $k_{\rm max} = 2\pi/\ell_{\rm th}$, which defines the thermalization length $\ell_{\rm th}$. Perturbation theory is controlled because all couplings in the Lagrangian (\ref{eq_action_main}) are power counting irrelevant, and therefore have small effects at low, hydrodynamic energy scales. Indeed, the appropriate dimensional analysis is set by the diffusive pole in (\ref{eq:bare}), so that $[\omega] = 2 [k]$ and $[D]=0$. It follows that $[\varphi_a] = [\varepsilon]= \frac{d}{2}[k]$ and hence the cubic couplings $[\lambda] = [\widetilde\lambda] = -\frac{d}{2}[k]$ are irrelevant. These dimensions suggest that the one loop corrections to tree-level diffusion, which are quadratic in coupling, will be of the schematic form $Dk^2(1+\lambda^2 k^d)$, as we verify below.

\begin{figure}[h]
\centerline{
\subfigure{\label{sfig_label1}
\includegraphics[width=0.22\linewidth, angle=0]{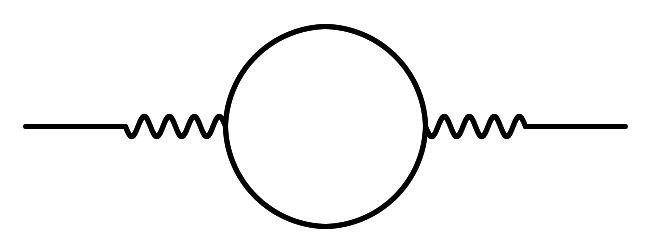}}
\subfigure{\label{sfig_label2}
\includegraphics[width=0.22\linewidth, angle=0]{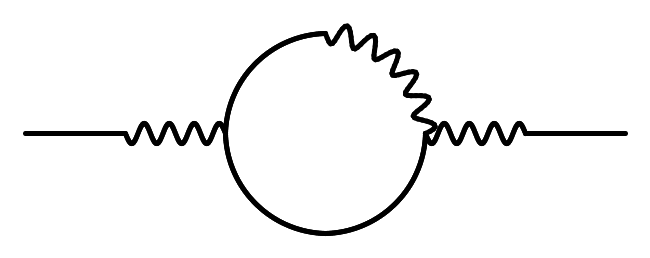}}
\subfigure{\label{sfig_label2}
\includegraphics[width=0.22\linewidth, angle=0]{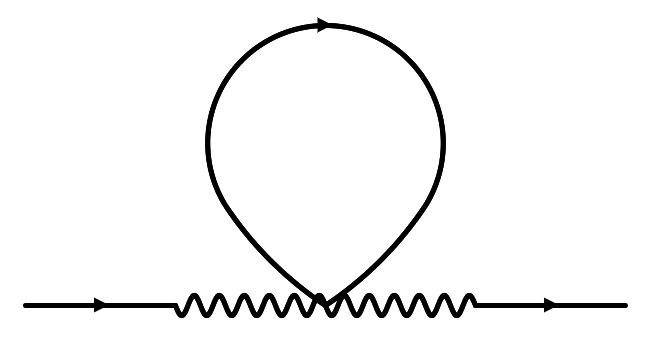}}
}
\centerline{
\subfigure{\label{sfig_label2}
\includegraphics[width=0.22\linewidth, angle=0]{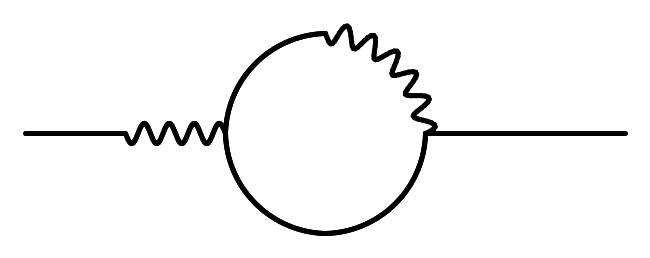}}
\subfigure{\label{sfig_label2}
\includegraphics[width=0.22\linewidth, angle=0]{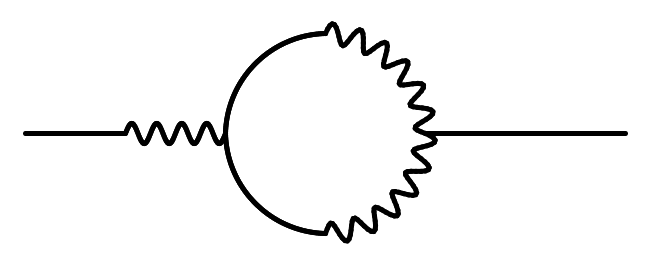}}
\subfigure{\label{sfig_label2}
\includegraphics[width=0.22\linewidth, angle=0]{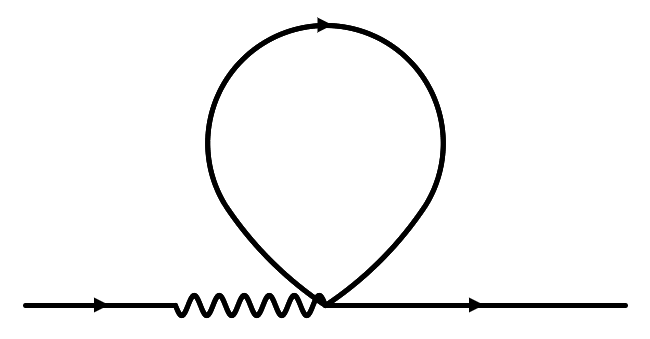}}
}\caption{\label{fig_loops}The one-loop diagrams contributing to $G_{\varepsilon \varepsilon}$. Solid lines denote the energy density field $\varepsilon$ and squiggly lines denote the auxiliary field $\varphi_a$.}
\end{figure}

\section{Results}

The diagrams shown in Fig. \ref{fig_loops} sum up to give the one loop retarded Green's function
\begin{equation}\label{eq_GR}
G^R_{\varepsilon \varepsilon}(\omega,k)
	= \frac{i \left(\kappa + \delta\kappa(\omega,k)\right) Tk^2}{\omega+ i D k^2 + \Sigma(\omega,k)}\, ,
\end{equation}
where both $\delta\kappa(\omega,k)$ and $\Sigma(\omega,k)$ receive analytic and non-analytic contributions. Separating these contributions as
\begin{equation}\label{eq_ana_nonana}
\delta \kappa(\omega,k)
	= \delta \kappa + \kappa_\star(\omega,k) \, , \qquad\qquad
\Sigma(\omega,k)
	= i\delta D k^2 + \Sigma_\star(\omega,k) \, ,
\end{equation}
one finds that the analytic pieces have the form
\begin{equation}\label{eq_res_ana}
\frac{\delta \kappa}{\kappa} = \frac{f_d}{c\, \ell_{\rm th}^d} \lambda_\kappa\, , \qquad\qquad
\frac{\delta D}{D} = \frac{f_d}{c\, \ell_{\rm th}^d} \lambda_D\, ,
\end{equation}
with $f_d = {\rm Vol} (B_{d})  =  2,\, {\pi},\, \frac{4\pi}{3}$ for spatial dimensions $d=1,2,3$, and where $\lambda_\kappa$, $\lambda_D$ are dimensionless effective couplings. Their explicit form will be given below. The non-analytic parts of \eqref{eq_ana_nonana} have the form
\begin{equation}\label{eq_res_nonana}
\kappa_\star(\omega,k)
	= f_\kappa(\omega,k) \alpha_d (\omega,k) \, , \qquad\qquad
\Sigma_\star(\omega,k)
	= k^2 f_\Sigma(\omega,k) \alpha_d (\omega,k) \, , 
\end{equation}
where $f_\kappa$, $f_\Sigma$ are analytic functions, shown below, that do not depend on dimension, and the non-analyticity is 
\begin{subequations}\label{eq_alpha}
\begin{align}
&&\alpha_1(\omega,k) &= \frac{1}{4} \left[k^2 - \frac{2i\omega}{D}\right]^{-1/2}\, , 
& (d=1)\\
&&\alpha_2(\omega,k) &= -\frac{1}{16\pi}\log \left[k^2 - \frac{2i\omega}{D}\right]\, , 
& (d=2)\\
&&\alpha_3(\omega,k) &= -\frac{1}{32\pi} \left[k^2 - \frac{2i\omega}{D}\right]^{1/2}\, .
& (d=3)
\end{align}
\end{subequations}
The effect of these non-analyticities is suppressed by powers of momenta and frequency appearing in $f_\kappa$, $f_\Sigma$, as we will see below.

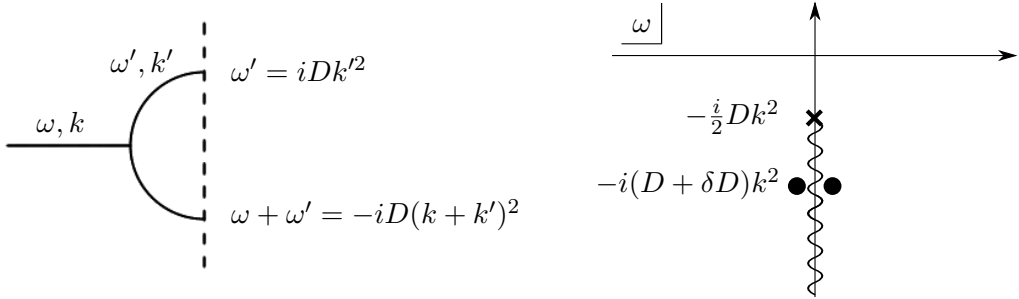
\begin{figure}
\centerline{
\subfigure{\label{sfig_label1}
\begin{overpic}[width=0.40\textwidth,tics=10]{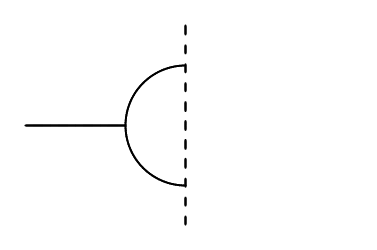}
	 \put (55,47) {$\omega'=iD k'^2$}
	 \put (55,16) {$\omega+\omega'=-iD (k+k')^2$}
	 \put (13,36) {$\omega,k$} 
	 \put (29,49) {$\omega',k'$} 
\end{overpic} 
}
\hspace{50pt}
\subfigure{\label{sfig_label2}
\begin{overpic}[width=0.35\textwidth,tics=10]{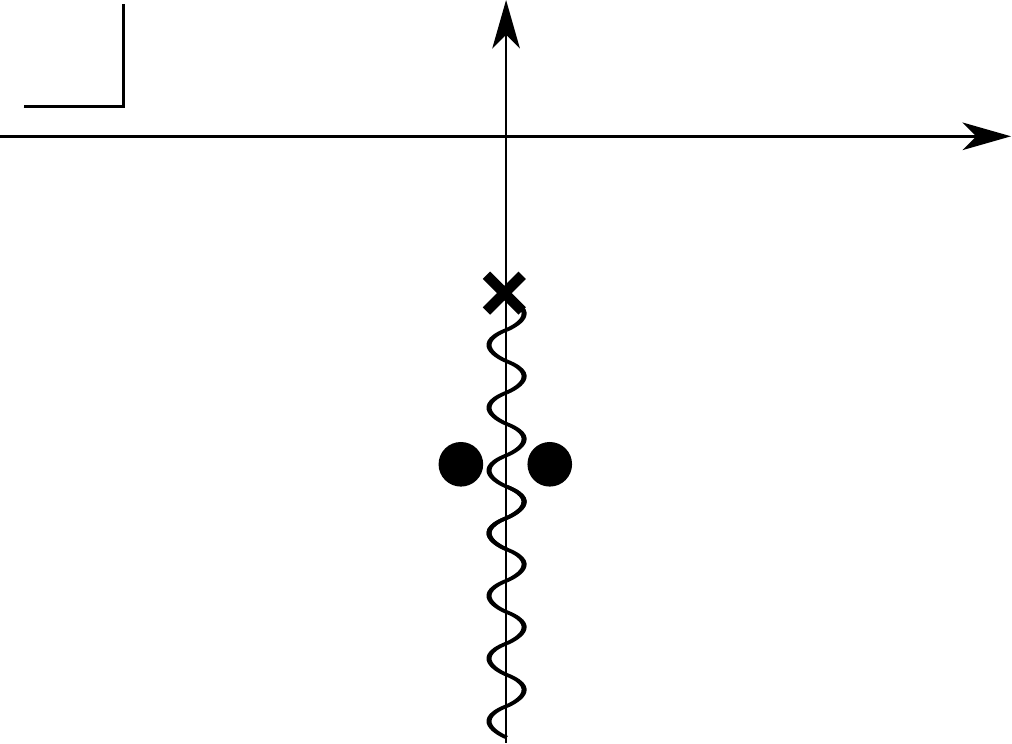}
	 \put (5,66) {$\omega$}
	 \put (18,43) {$ - \frac{i}{2}D k^2$} 
	 \put (-4,26) {$ - i(D+\delta D)k^2$} 
\end{overpic}
}
}
\caption{\label{fig_non_ana} On-shell condition for the two internal legs (left), and analytic structure of the retarded Green's function $G^R_{\varepsilon\varepsilon}(\omega,k)$ at one loop (right). In imposing the on-shell condition, it is important to consider two poles in opposite halves of the complex $\omega'$ plane, otherwise the loop contribution vanishes. The pole in the upper half plane arises from an advanced Green's function in the loop: $G^A = \left(G^R\right)^*$.}
\end{figure}

The retarded Green's function is analytic in the upper-half frequency plane, as required by causality. 
The interactions have induced a branch point at $\omega = -\frac{i}{2} D k^2$. Moreover, the diffusive pole is split into two poles with small real parts $\omega = -i(D+\delta D)k^2 \pm O(k^2 |k|^{d}).$\footnote{Additional poles in \eqref{eq_GR} are outside the validity of the resummation. The non-analytic corrections to diffusion are seen to be more important than the $O(k^4)$ higher derivative corrections to the diffusion equation in $d=1$ and also in $d=2$, where the dispersion receives an additional imaginary part $O(k^4 \log k)$. 
}  The location of the branch point can be understood from simple kinematics, by putting both internal legs on-shell (in either the retarded or advanced Green's functions) as in Fig.~\ref{fig_non_ana}. The frequencies $\omega$ for which the on-shell condition is satisfied form a half-line in the complex plane parametrized by the loop momentum $k'$, where the Green's function has a branch cut. The branch point is located at the smallest frequency $\omega$ (in magnitude) that can satisfy the on-shell conditions:
\begin{equation}
\omega_\star
	= -i D \min_{k'} \left[ k^2 + 2k\cdot k' + 2 k'^2\right]
	= -\frac{i}{2}Dk^2\, .
\end{equation}

In previous treatments, similar physics to what we have just described was found in the coupled diffusion of two modes \cite{PhysRevB.73.035113,Kovtun:2014nsa}. Due to the absence of a systematic formalism for hydrodynamic fluctuations at that time, those works did not account --- among other things --- for interactions that are quadratic in the auxiliary field (in particular $\widetilde\lambda$), nor the quartic terms $\lambda'$ and $\widetilde\lambda'$. While the systematic approach modifies the results for two coupled modes, see appendix \ref{app_2n}, the most qualitative difference is seen for diffusion of a single conserved density. We have found that renormalization of the diffusion constant and conductivity occurs even in this case, 
controlled by the effective couplings in \eqref{eq_res_ana}
\begin{equation}\label{eq:ll}
\lambda_\kappa = \frac{c^2 T^2 }{D}\tilde\lambda'\, , \qquad\qquad
	\lambda_D = - \frac{c^2 T^2}{2D^2} \left[\lambda(\lambda+\tilde \lambda) + 2\lambda' D\right]\,,
\end{equation}
which are not sign-definite in general.

Furthermore, we have also found non-analytic corrections to the Green's function even with a single diffusing mode. These are not as strong as those arising with two modes, as we now explain.
The functions $f_\kappa,\,f_\Sigma$ appearing in the non-analytic contributions \eqref{eq_res_nonana} are
\begin{equation}
f_\kappa(\omega,k)
	= \frac{cT^2}{D^2} k^2 \lambda\tilde \lambda\, , \qquad\qquad
f_\Sigma(\omega,k)
	=\frac{cT^2}{D^2}\left[\omega \lambda(\lambda+\tilde \lambda) + iDk^2 \lambda\tilde \lambda\right]\, .
\end{equation}
While $f_\Sigma$ has both the $O(\omega)$ and $O(k^2)$ terms expected at this order in the derivative expansion, $f_{\kappa}$ only has an $O(k^2)$ term.
The subsequent suppression of $f_{\kappa}$ as $k \to 0$ implies that the optical conductivity does not receive non-analytic corrections, and is instead constant in the hydrodynamic regime
\begin{equation}
T\kappa(\omega)
	\equiv \lim_{k\to 0} \frac{\omega}{k^2}\Im  G^R_{\varepsilon \varepsilon}(\omega,k)
	= T(\kappa + \delta \kappa)\, .
\end{equation}
This result can be contrasted with the case of two interacting diffusive densities, wherein the optical conductivity receives a non-analytic fluctuation correction \cite{DESCHEPPER19741,PhysRevB.73.035113}. We revisit this case in appendix \ref{app_2n}, where the analytic structure is discussed in the light of a systematic inclusion of fluctuation effects. We find a non-analytic correction to the optical conductivity of the form (we use $\sigma$ to denote a generic conductivity)
\begin{equation}
\delta\sigma(\omega,k)
	\sim \omega\, \alpha_d(\omega,k) + \cdots\, ,
\end{equation}
where $\alpha_d$ is as in (\ref{eq_alpha}) and $\cdots$ denote terms that are further $k^2$ suppressed. In particular, the correction in $d=2$ is $\delta\sigma(\omega) \sim \omega\log\omega$.

\section{Discussion and Applications}

Strong renormalization of the transport parameters due to hydrodynamic fluctuations occurs if the ratios in (\ref{eq_res_ana}) are large. When the dimensionless couplings are order unity, $\lambda_\kappa\sim \lambda_D \sim 1$,\footnote{Using equations (\ref{eq:ll}) and (\ref{eq_couplings}), this holds in any scaling regime where $\kappa \sim T^a$ and $c \sim T^b$.} the strength of fluctuations is controlled by the specific heat per `thermal volume' $c\ell_\text{th}^d$. This quantity can be thought of as the number of degrees of freedom in the smallest volume that can reach local thermodynamic equilibrium. When there are many degrees of freedom in a thermal volume, fluctuation effects are small. One might further expect that a sufficient number of degrees of freedom are necessary in order for a region to locally thermalize, and hence $c\ell_\text{th}^d \gtrsim 1$. Indeed, such a bound has been established in the presence of operators with microscopic positivity properties, using the eigenstate thermalization hypothesis \cite{Delacretaz:2018cfk,ETHnew}. Thus hydrodynamic fluctuation corrections to thermal transport parameters are expected to be at most comparable to the bare values.

If microscopic interactions are weak then $\ell_\text{th} \sim \ell_\text{mfp}$, the (inelastic) quasiparticle mean free path, will be large. Fluctuation corrections to transport are therefore small in weakly interacting systems. In contrast, in strongly correlated systems $\ell_\text{th}$ can become very short and fluctuations may be important. For example, at high temperatures in a lattice model, with order one degrees of freedom per unit cell,
the bound mentioned above is only saturated when 
$\ell_\text{th} \sim a$, the lattice spacing. This roughly coincides with the `minimal' mean free path for thermal transport by well-defined phonons in insulators \cite{SLACK19791, PhysRevB.49.9073}. In strongly correlated regimes, however, the notion of a mean free path is likely not a useful concept. Recent measurements of thermal diffusivity in cuprates \cite{Zhang:2016ofh,2018arXiv180807564Z} and perovskites \cite{doi:10.1063/1.3371815,PhysRevLett.120.125901,AhBh} suggest (assuming that the microscopic sound speed is the relevant velocity) that a transport lengthscale reaches and possibly surpasses the lattice spacing at high temperatures. The specific heat in these materials is roughly $c a^2 \sim 40$ and $c a^3 \sim 15$, respectively. It may be interesting to look for signatures of diffusive fluctuations in the thermal transport of these systems.

Fluctuation effects can also become important for transport close to a thermal phase transition. The thermalization length diverges as $\ell_\text{th} \sim \tau^{-\nu}$ as the reduced temperature $\tau \to 0$, while the specific heat scales as $c \sim \tau^{2-\alpha}$. It follows that $\delta D/D \sim \tau^{\alpha + d\nu-2} \sim 1$ if hyperscaling is obeyed, so fluctuations are important in that case. Above the upper critical dimension hyperscaling is violated and fluctuations are small.  A more sophisticated discussion must include fluctuations of the order parameter in the analysis \cite{RevModPhys.49.435}.

Transport lengthscales approaching or exceeding the lattice spacing are also seen in `bad metals' \cite{PhysRevLett.74.3253,RevModPhys.75.1085,hussey}. All of our expressions above are easily adapted to describe the diffusion of a single conserved $U(1)$ charge, instead of heat.\footnote{If thermoelectric effects are strong, one should instead work with coupled heat and charge diffusion, as in appendix \ref{app_2n}.} Particle-hole symmetry should be broken, typically by a background charge density, otherwise many terms we have considered are forced to be zero.
The correction to the dc electrical conductivity $\sigma$, for example, is found to be
\begin{equation}\label{eq:sigma}
\frac{\delta \sigma}{\sigma}
	= \frac{f_d}{\ell_{\rm th}^d} \frac{T}{\chi \mu^2} \lambda_\sigma\, .
\end{equation}
Here $\chi$ is the charge susceptibility and $\mu$ the chemical potential.
In the definition of the couplings in (\ref{eq_action_main}) in terms of the thermodynamic derivatives (\ref{eq_couplings_raw}), one replaces $T \to \mu$.

Condensed matter systems --- including most bad metals --- are typically at degenerate temperatures $T < E_F$, below the Fermi energy. At these temperatures $\chi \sim k_F^d/E_F$ and $\mu \sim E_F$. Here $k_F$ is the Fermi momentum. The contribution (\ref{eq:sigma}) of fluctuations to the conductivity is therefore small, even when the thermalization length becomes of order $\ell_\text{th} \sim a \sim 1/k_F$. This is the shortest length consistent with local thermalization \cite{ETHnew}. In contrast, at high temperatures where fermions are non-degenerate, $\chi \sim 1/(T a^d)$. If the total charge is held fixed, then $\mu \sim T$. It follows that as the thermalization length becomes short, of order $\ell_\text{th} \sim a$, fluctuation corrections to the conductivity are order one. Diffusive transport by strongly correlated but non-degenerate fermions has recently been probed in an ultracold atom realization of the Hubbard model \cite{2018arXiv180209456B}, and earlier in e.g. \cite{Schneider2012}, as well as in numerics \cite{2018arXiv180308054M, 2018arXiv180608346H}. Indeed, $\ell_\text{th}$ is found to saturate around the lattice scale at high temperatures, and so fluctuation effects may be important.

Finally, diffusion with a short thermalization length has also been seen in spin transport in strongly interacting ultracold atoms in a trap \cite{2018arXiv180505354E}. The formulae we have developed can be applied directly to longitudinal spin diffusion in a magnetic field (to break spin reversal symmetry) or to transverse spin diffusion without a magnetic field or spontaneous magnetization (so that isotropy prevents mixing of the two transverse modes).
At temperatures $T \lesssim E_F$, with electrons on the verge of becoming non-degenerate, the thermalization length is found to be $\ell_\text{th} \sim 1/k_F$ (there is no lattice scale in these experiments). Diffusive fluctuations may therefore again be important for transport.

In summary, long wavelength fluctuations about diffusive dynamics may be relevant in condensed matter and cold atom systems of widespread interest. We have seen that a systematic derivation of these effects leads to different results than previous, more phenomenological, approaches. For this reason, it will be important to revisit the computation of fluctuations in relativistic hydrodynamics \cite{Kovtun:2011np, Stephanov:2017ghc}, which includes a sound mode in addition to transverse momentum diffusion. Fluctuations in relativistic hydrodynamics may have direct consequences for the quark-gluon plasma.

\section*{Acknowledgments} 
 
We would like to thank Erez Berg, Debanjan Chowdhury, Paolo Glorioso and Andrew Lucas for illuminating discussions. SAH and LVD are partially supported by the US Department of Energy Office of Science under Award Number DE-SC0018134. XCL is supported by the Knut and Alice Wallenberg Foundation.

\pagebreak

\bibliographystyle{ourbst}
\bibliography{diff}{}

\newpage

\appendix

\section{Effective action formalism for hydrodynamic fluctuations}\label{app_Hong}
We are interested in the energy density correlation functions in an interacting non-integrable quantum system at finite temperature
\begin{equation}\label{eq_corr_app}
\langle \varepsilon (t, x) \varepsilon (t', x') \cdots\rangle_\beta
	\equiv \Tr \Bigl(\rho_{\beta}\,  \varepsilon (t,x) \varepsilon(t',x')\cdots \Bigr)\, ,
\end{equation}
where $\rho_\beta = e^{-\beta H}/\Tr e^{-\beta H}$. We assume that energy is the only conserved quantity.

We will address this problem using the effective hydrodynamic action formalism of Ref.~\cite{Crossley:2015evo}, which we review below. Effective field theories for dissipative hydrodynamics have 
seen a fast development recently \cite{Endlich:2012vt,Grozdanov:2013dba,Haehl:2015pja,Crossley:2015evo} (see \cite{Glorioso:2018wxw} for a recent review). The resulting formalism makes transparent the physical principles that constrain the dynamics of conserved densities and their interactions with microscopic excitations (or `noise'), and offers a systematic perturbative expansion to study fluctuation corrections to hydrodynamic correlators. It is more general than conventional stochastic approaches \cite{POMEAU197563, PhysRevLett.56.889} which it reproduces in certain limits.

We review the path integral formulation for correlators of the form \eqref{eq_corr_app} in \ref{app_ctp}, and construct the effective action for energy density fluctuations in \ref{app_L2} and \ref{app_L3}. This is a special case of the more general formalism of Ref.~\cite{Crossley:2015evo} which we present here for completeness. In appendix \ref{app_loops} we use this formalism to compute the one loop correction to the diffusive Green's function $G_{\varepsilon \varepsilon}$.

\subsection{Generating functional on a closed time path}\label{app_ctp}

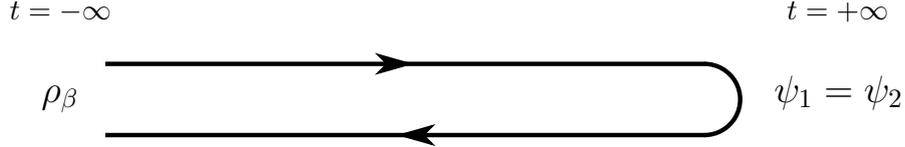
\begin{figure}[h]
\vspace{30pt}
\centerline{
	\begin{overpic}[width=0.55\textwidth,tics=10]{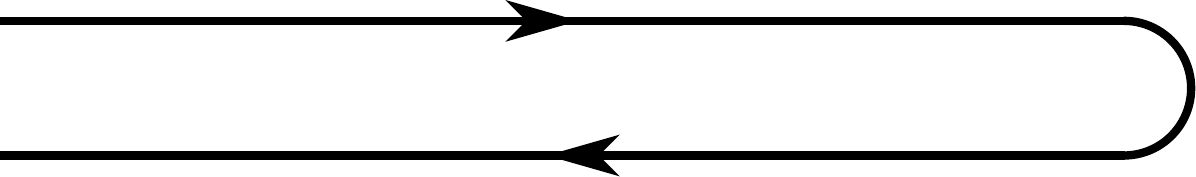}
	 \put (-15,20) {$t= -\infty$} 
	 \put (107,20) {$t=+\infty$} 
	 \put (-10,7) {{\Large $\rho_{\beta}$}} 
	 \put (105,7) {{\Large $\psi_1=\psi_2$}} 
\end{overpic} 
}
\caption{\label{fig_SK} Closed time path.}
\end{figure}

Correlation functions of the form \eqref{eq_corr} can be studied by coupling the conserved energy current $j^\mu$ to background gauge fields
\begin{equation}\label{eq_Z_1}
Z[A^1_\mu,A^2_\mu,]
	\equiv \Tr \left( U[A_1]\rho_\beta U[A_2]^\dagger\right)\, ,
\end{equation}
where $U$ is the time evolution operator from $t=-\infty$ to $+\infty$. By taking derivatives with respect to the appropriate sources $A^1,\,  A^2$ one can compute the correlators in \eqref{eq_corr} with various time orderings. In terms of the microscopic fields and action $S[\psi]$, the trace in \eqref{eq_Z_1} can be represented as a path integral on a closed time path%
	\footnote{Out of time order correlators can also be computed in this framework but the contour needs to have two more switchbacks \cite{Blake:2017ris}.}
\begin{equation}\label{eq_Z_micro}
Z[A^1_\mu, A^2_\mu]
	= \int_{} D\psi_1 D\psi_2 \, e^{iS[\psi_1,A_1] - iS[\psi_2,A_2]}
	\, , 
\end{equation}
where $S[\psi,A] = S[\psi] + \int d^{d+1}x\, A_\mu j^\mu$ and $\psi_1$ ($\psi_2$) denote the fields on the upper (lower) leg of the path shown in Fig.~ \ref{fig_SK} -- they are identified at $t= + \infty$ because of the trace.
The generating functional satisfies several important properties which will be useful to constrain the effective hydrodynamic action later:
\begin{subequations}\label{eq_Z_identities}
\begin{align}
 && Z[A_\mu , A_\mu] \label{seq_Z1}
	&= 1\, , \\
\hbox{reflection:} && Z[A_\mu^1,A_\mu^2] \label{seq_Z2}
	&= Z^*[A_\mu^2,A_\mu^1]\, ,  \\
\hbox{gauge invariance:} && Z[A_\mu^1, A_\mu^2] \label{seq_Z3}
	&= Z[A_\mu^1 + \d_\mu\lambda^1, A_\mu^2+ \d_\mu\lambda^2]\, ,  \\
\hbox{KMS condition:} && Z[A_\mu^1,A_\mu^2] \label{seq_Z4}
	&= Z[A_\mu^1 (-t,x_{PT}),A_\mu^2(-t- i\beta,x_{PT})]\, , &&
\end{align}
\end{subequations}
all of which follow directly from \eqref{eq_Z_1}. In writing the last identity we have further assumed invariance under PT symmetry.

\subsection{Quadratic action}\label{app_L2}

The generating function $Z[A_\mu^1,A_\mu^2]$ is non-local in the background fields, because massless hydrodynamic excitations have been integrated out. A central assumption of \cite{Crossley:2015evo} is that they can be integrated back in with the St\"uckelberg trick, i.e.~that there exists a local effective action $I$ satisfying
\begin{equation}
Z[A_\mu^1 , A_\mu^2]
	= \int D\varphi_1D\varphi_2 \, e^{i I[B^1_\mu,B^2_\mu]}\, , \label{eq:ZI}
\end{equation}
where $B_{\mu} = A_\mu + \d_\mu \varphi$. This expression automatically satisfies gauge invariance \eqref{seq_Z3}. Notice that unlike in the formulation in terms of microscopic fields \eqref{eq_Z_micro}, the emergent degrees of freedom $\varphi_{1,2}$ on both legs of the closed time path can interact \cite{Grozdanov:2013dba}. We will construct the effective action $I[B^1_\mu,B^2_\mu]$ subject to the conditions \eqref{eq_Z_identities}. For simplicity, we will only write expressions for the effective action in the absence of sources ($A_\mu = 0$), in a gradient expansion and perturbatively in fields. It will be convenient to introduce `classical' and `auxiliary' fields as
\begin{equation}
\varphi_r = 
	\frac{1}{2} (\varphi_1 + \varphi_2)\, , \qquad \qquad
\varphi_a = 
	 \varphi_1 - \varphi_2\, .
\end{equation}
We will see below that $\dot\varphi_r$ is related to the energy density $\varepsilon$, and $\varphi_a$ to the noise field commonly used in KPZ-like equations. For this reason, it is convenient to adopt from the start a derivative counting scheme such that $\dot\varphi_r\sim \varphi_a$. Moreover, because only $\dot \varphi_r$ is physical, one can choose to further impose upon $I$ a time-independent local shift symmetry
\begin{equation}\label{eq_shift}
\varphi_r \to \varphi_r + \lambda(\vec x) \,.
\end{equation}
The physics of this symmetry will be discussed at the end of this section.

We start with the quadratic action $I_2 = \int d^{d+1}x \, \mathcal L_2$. The condition \eqref{seq_Z1} implies $Z[A_{a\mu}=0,A_{r\mu}]=0$, and hence any correlator involving only auxiliary fields must vanish (since $A_1j_1 + A_2 j_2 = A_aj_r + A_r j_a$, correlators of $j_a$ are obtained from $\delta/\delta A_{r}$). In particular $\langle \varphi_a \varphi_a\rangle=0$ requires that the quadratic action start at linear order in $\varphi_a$. The most general isotropic action to leading order in derivatives is then%
	\footnote{In two spatial dimensions one can also add a parity-breaking term $\kappa_{\rm H} \epsilon^{ij}B_{i}^a\dot B_{j}^r$ which leads to a Hall response. Since this term is a total derivative once the sources are removed ($A\to 0$) it does not enter in density correlators; we will therefore ignore such terms here. They do however enter in the transverse part of current correlators, see e.g.~the appendix of \cite{PhysRevB.97.220506}. The same statement also applies to interactions, suggesting that the Hall sector does not get renormalized at one loop to leading order in derivatives.}
\begin{equation}\label{eq_L2_1}
\beta\mathcal L_2
	= 
	c \dot \varphi_r  \dot \varphi_a
	 + \kappa \dot \varphi_r \nabla^2\varphi_a 
	 + i T \widetilde\kappa (\nabla\varphi_a)^2 
	 + \cdots\, ,
\end{equation}
where all the coefficients $c,\,\kappa,\, \widetilde \kappa$ are real and the factors of $i$ are required by reflection symmetry \eqref{seq_Z2}. Our derivative expansion with $\dot\varphi_r\sim \varphi_a$ implies for example that $\dot \varphi_a^2$ is subleading compared to the $c$ term above. The shift symmetry \eqref{eq_shift} requires a time derivative on the $\kappa$ term. Finally, we need to impose the KMS symmetry \eqref{seq_Z4}. This is done in appendix \ref{sec_KMS} and fixes
\begin{equation}
\widetilde \kappa = \kappa\, .
\end{equation}
The final quadratic action is therefore
\begin{equation}
\mathcal L_2
	= i T^2\kappa (\nabla \varphi_a)^2 - \varphi_a \left(\dot \varepsilon - \frac{\kappa}{c} \nabla^2 \varepsilon\right)\, ,
\end{equation}
where we identified%
	\footnote{Restoring the sources $\dot\varphi_a\to A_{a0} + \dot\varphi_a$, the density is given by $j^0_r = \frac{\delta I}{\delta A_{a0}}= c T \dot\varphi_r\equiv \varepsilon$. } 
$\varepsilon=c T \dot \varphi_r$. Correlators of the physical density in this theory are obtained by inverting the kinetic term, and give the symmetric Green's function
\begin{equation}
G_{\varepsilon \varepsilon}(\omega,k)
	= \langle \varepsilon \varepsilon\rangle_\beta (\omega,k)
	= \frac{2\kappa T^2 k^2}{\omega^2 + (D k^2)^2}\, ,
\end{equation}
with diffusivity $D = \kappa/c$.

We now briefly comment on the shift symmetry \eqref{eq_shift}. Lifting it would allow a term more relevant than $\kappa$ in the effective action \eqref{eq_L2_1}
\begin{equation}	
\kappa \dot\varphi_r \nabla^2 \varphi_a \quad \to\quad 
c v_s^2 \varphi_r \nabla^2 \varphi_a \, .
\end{equation}
Instead of diffusive Green's functions, this system would exhibit a sound pole $\omega = \pm v_s k + \ldots$, as expected if the symmetry were spontaneously broken (in the case of an internal $U(1)$ symmetry, this would describe a superfluid). Since we assume time-translation symmetry is preserved, we choose to impose the shift symmetry to the effective action. See \cite{Crossley:2015evo} for further discussions on this symmetry.

\subsection{Interactions}\label{app_L3}

The most general cubic action that satisfies the constraints discussed in the previous section is given by
\begin{equation}
\beta\mathcal L_3
	= iT \widetilde\kappa_1 (\nabla \varphi_a)^2 \dot \varphi_r + \frac{1}{2}\kappa_1 (\nabla^2 \varphi_a) \dot \varphi_r^2 + \frac{1}{2} c_1 \dot\varphi_a \dot\varphi_r^2 + \cdots \, ,
\end{equation}
where $ \widetilde\kappa_1,\, \kappa_1,\, c_1$ are real coefficients. The derivative counting scheme $\dot \varphi_r \sim \varphi_a$ and $\d_t \sim \nabla^2$ implies for example that $\dot\varphi_a^2 \dot\varphi_r$ and any $O(\varphi_a^3)$ term would be subleading compared to the $c_1$ term. The interactions must also be at least linear in $\varphi_a$. Indeed, an interaction $\sim \varphi_r^3$ can easily be seen to generate a non-zero correlator $\langle \varphi_a \varphi_a \rangle\neq 0$ at one-loop (and $\langle \varphi_a \varphi_a\varphi_a \rangle\neq 0$ at tree-level), violating \eqref{seq_Z1}.

The quartic action is similarly found to be
\begin{equation}
\beta\mathcal L_4
	= \frac{1}{2}iT \widetilde\kappa_2 (\nabla \varphi_a)^2 \dot \varphi_r^2 - \frac{1}{12}\kappa_2 (\nabla^2 \varphi_a) \dot \varphi_r^3 + \frac{1}{6} c_2 \dot\varphi_a \dot\varphi_r^3 +  \ldots\, ,
\end{equation}
where again $O(\varphi_a^3)$ and $O(\varphi_a^4)$ terms are suppressed in derivatives%
	\footnote{This suppression of higher order terms in the auxiliary field was found using a different approach in \cite{Banks:2018aob}.}.
Since we are interested in one-loop corrections to the propagator, we will not need higher order interactions. KMS conditions are imposed in section \ref{sec_KMS} and lead to
\begin{equation}
\widetilde\kappa_1 = \kappa_1 \, , \qquad \qquad
\widetilde\kappa_2 = \kappa_2 \, .
\end{equation}

The couplings are related to derivatives of the parameters appearing in the quadratic action with respect to temperature. This is seen as follows. Firstly, the energy density can be identified by reintroducing the source $\dot\varphi_a \to A_{a0} + \dot\varphi_a$ :
\begin{equation}\label{eq_ch_var}
\begin{split}
\varepsilon = j^0_r = \frac{\delta I}{\delta A_{a0}}
	&= cT \dot \varphi_r + \frac12 c_1 T\dot\varphi_r^2
	+ \frac16 c_2 T \dot\varphi_r^3 + \cdots \, \\
	&= c \delta T + \frac12 c_1 \frac{\delta T^2}{T}
	+ \frac16 c_2 \frac{\delta T^3}{T^2} + \cdots \, ,\\
\end{split}
\end{equation}
where in the second line we identified the source for energy fluctuations $\dot \varphi_r \equiv \delta T/T$.%
    \footnote{In the case of charge conservation, the local chemical potential is naturally defined as $\mu(x) = B_0(x) = \mu + \dot \varphi_r(x)$, where $\mu = A_0$ is the equilibrium chemical potential. This implies $\delta \mu(x) = \dot\varphi_r(x)$. For energy conservation, $\mu$ is replaced by the reduced temperature, which sources energy density \cite{PhysRev.135.A1505}.}
Taking the expectation value (where the normal ordering prescription implies e.g.~$\langle \delta T^2 \rangle = \langle \delta T\rangle^2$) we see that
\begin{equation}
c_n 
    = T^{n} \frac{\d^{n+1} \varepsilon}{\d T^{n+1}}
    = T^n \frac{\d^n c}{\d T^n}\,.
\end{equation}
Similarly (note that the auxiliary fields do not have expectation values)
\begin{equation}
\begin{split}
j^i_r = \frac{\delta I}{\delta A_{ai}}
	&= \left[\kappa + \kappa_1 \frac{\delta T}{T} + \frac12\kappa_2 \frac{\delta T^2}{T^2}+\ldots\right]\frac{\nabla T}{T}\, ,
\end{split}
\end{equation}
which implies
\begin{equation}
\kappa_n = T^n \frac{\d^n \kappa}{\d T^n}\, .
\end{equation}
Therefore, making the change of variable \eqref{eq_ch_var} in the action leads to 
\begin{equation}\label{eq_action}
\mathcal L_{\rm int} = 
\nabla^2\varphi_a\left[\frac{1}{2}\lambda\varepsilon^2 + \frac13\lambda'\varepsilon^3\right] +
icT^2(\nabla\varphi_a)^2 \left[\widetilde\lambda\varepsilon + \widetilde\lambda'\varepsilon^2\right]
	+ \cdots \, ,
\end{equation}
where the four couplings are given by
\begin{equation}\label{eq_couplings}
\lambda = \frac{D_1}{cT}\, , \qquad
\widetilde\lambda = \frac{\kappa_1}{c^2 T}\, , \qquad
\lambda' = - \frac{3 c_2 D + 8 c_1 D_1 + c D_2}{4c^3T^2 }\, , \qquad
\widetilde\lambda' = \frac{\kappa_2 c - \kappa_1 c_1}{2c^4T^2}\, ,
\end{equation}
with $D_n \equiv T^n \d^n_T D$.

\subsection{KMS condition}\label{sec_KMS}

At the classical level, a sufficient condition for invariance under the KMS condition \eqref{seq_Z4} is that $I[ B_\mu^1,B_\mu^2]$ in (\ref{eq:ZI}) be invariant, with $B$ transforming in the same way as the background $A$. 
One way to ensure that this condition holds beyond tree-level is to introduce ghost fields \cite{Crossley:2015evo}. It can be shown that any ghost loop contribution to bosonic correlation functions vanishes \cite{Gao:2018bxz} (see also \cite{Jensen:2018hse}).\footnote{The essential fact is that the propagator of the ghost fields is a retarded Green's function whose poles are therefore constrained to lie in the lower half of the complex plane.} For this reason ghost fields can be ignored for our purposes, and we focus on the classical action.

Consider first the quadratic action. Only mixing terms $B_1 B_2$ are affected since the two fields will be evaluated at different times; these are
\begin{equation}
\beta \mathcal L_{\rm mix}
	= -i 2T \widetilde\kappa B_{1i}B_{2i} + \kappa B_{1i} \dot B_{2i}
	\, , \quad \hbox{or} \quad
\beta\mathcal L_{\rm mix} = 
	-i\left(2T \widetilde \kappa - \omega \kappa\right)B_1^{\omega,k} B_2^{-\omega,-k}
\end{equation}
in momentum space. If following \eqref{seq_Z4} time is translated by $i\beta$ in $B_2$ before Fourier transforming, we obtain 
\begin{equation}
\beta\mathcal L_{\rm mix}
	=-i\left(2T \widetilde \kappa + \omega \kappa\right) e^{-\beta \omega}B_1^{\omega,k} B_2^{-\omega,-k}\, .
\end{equation}
Equating these two expressions leads to the constraint
\begin{equation}
\widetilde \kappa 
	= \kappa \frac{\beta\omega}{2} \coth\frac{\beta\omega}{2}
	= \kappa + O(\beta\omega)\, ,
\end{equation}
(the $\omega$ dependence of $\widetilde\kappa$ are higher derivative corrections in the action $\widetilde\kappa\to \widetilde\kappa + \widetilde\kappa'\d_t + \ldots$ which we have already dropped). Setting $\kappa = \widetilde\kappa$ therefore guarantees that the quadratic action satisfies the KMS condition up to higher derivative terms.

These constraints can similarly be imposed on the cubic and quartic terms. The derivations are however quite lengthy, so we refer the reader to Refs.~\cite{Wang:1998wg,Crossley:2015evo} and simply quote and apply their results. For a cubic effective action of the form
\begin{equation}
I_3[B^1,B^2] = 
	\frac{1}{2} G_{rra}^{\mu_1\mu_2\mu_3} B_{a\mu_1}B_{a\mu_2}B_{r\mu_3}
	+\frac{1}{2} G_{raa}^{\mu_1\mu_2\mu_3} B_{a\mu_1}B_{r\mu_2}B_{r\mu_3} \,,
\end{equation}
the constraint is (in the limit $\omega\ll T$)
\begin{equation}
\beta G_{rra} = -\left(\frac{1}{\omega_1} + \frac{1}{\omega_2}\right)G_{aar}^* + \frac{1}{\omega_1} G_{ara} + \frac{1}{\omega_2}G_{raa}\, .
\end{equation}
Applying this to the cubic action with
\begin{subequations}
\begin{align}
G_{rra}^{\mu_1\mu_2\mu_3}
	&= 2T \widetilde\kappa_1 \delta^{\mu_1}_i \delta^{\mu_2}_i \delta^{\mu_3}_0 \,,\\
G_{raa}^{\mu_1\mu_2\mu_3}
	&= -\kappa_1 \delta^{\mu_1}_i \left(\delta^{\mu_2}_0 \delta^{\mu_3}_i(-i\omega_3)+\delta^{\mu_3}_0 \delta^{\mu_2}_i(-i\omega_2)\right)\,, \, 
\end{align}
\end{subequations}
leads to $\kappa_1 = \widetilde\kappa_1$. For a quartic action
\begin{equation}
I_4[B^1,B^2] = 
	\frac{1}{4} G_{rraa}^{\mu_1\mu_2\mu_3\mu_4} B_{a\mu_1}B_{a\mu_2}B_{r\mu_3}B_{r\mu_4}
	+\frac{1}{6} G_{raaa}^{\mu_1\mu_2\mu_3\mu_4} B_{a\mu_1}B_{r\mu_2}B_{r\mu_3}B_{r\mu_4} \,,
\end{equation}
the constraint is
\begin{equation}
-\frac{\beta}{2}\left(\omega_1\omega_2 G_{rraa}+\omega_3\omega_4 G_{aarr}^*\right)
	= \omega_1 G_{raaa}+\omega_2 G_{araa}+ \omega_3 G_{aara}^*+\omega_4 G_{aaar}^*\, .
\end{equation}
Applying this to the quartic action with
\begin{subequations}
\begin{align}
G_{rraa}^{\mu_1\mu_2\mu_3\mu_4} 
	&= 2i T \tilde \kappa_2 \delta_{\mu_1}^i\delta_{\mu_2}^i \delta_{\mu_3}^0 \delta_{\mu_4}^0\, ,\\
G_{raaa}^{\mu_1\mu_2\mu_3\mu_4} 
	&=\frac{\kappa_2}{2} \delta_{\mu_1}^i \left(\delta_{\mu_2}^0 \delta_{\mu_3}^0 \delta_{\mu_4}^i (-i\omega_4) + (4 \leftrightarrow 3) + (4 \leftrightarrow 2)\right)\, ,
\end{align}
\end{subequations}
leads to $\kappa_2 = \widetilde\kappa_2$. The coefficents $c,\, c_1,\, c_2$ are not constrained by the KMS condition.

\section{One-loop correction to diffusion}\label{app_loops}

We write down the full action up to quartic order in fields for convenience
\begin{equation}
\mathcal L
	= iT^2\kappa (\nabla \varphi_a)^2 - \varphi_a \left(\dot \varepsilon -
	 D \nabla^2 \varepsilon\right) 
	+ \nabla^2\varphi_a\left[\frac12 \lambda\varepsilon^2 + \frac13\lambda'\varepsilon^3\right] +
	icT^2(\nabla\varphi_a)^2 \left[\widetilde\lambda\varepsilon + \widetilde\lambda'\varepsilon^2\right]
	+ \cdots \, ,
\end{equation}
with couplings given by \eqref{eq_couplings}.
The free propagators are 
\begin{equation}
G_{\varepsilon \varepsilon}^0 = \frac{2\kappa T^2 k^2}{\omega^2 + (D k^2)^2}\, , \qquad
G^0_{\varepsilon\varphi_a} = \frac{-1}{\omega+iD k^2}\, , \qquad
G^0_{\varphi_a \varepsilon} = \frac{1}{\omega-iD k^2}\, , \qquad
G^0_{\varphi_a\varphi_a} = 0\, ,
\end{equation}
The one-loop renormalized Green's function can be found by expanding the interaction and performing Wick contractions
\begin{equation}
\langle \varepsilon \varepsilon\rangle
	= \langle \varepsilon \varepsilon\rangle_0 + i \langle \varepsilon S_{\rm int} \varepsilon\rangle_0- \frac{1}{2} \langle \varepsilon S_{\rm int}^2 \varepsilon\rangle_0 + \cdots\, . \label{eq:ee}
\end{equation}

We will parametrize the corrections to the Green's function as a self-energy $\Sigma(\omega,k)$ and a numerator $C(\omega,k)$. This distinction corresponds to diagrams contributing to (\ref{eq:ee}) that resum and diagrams that do not, respectively.
The numerator $C(p)$ is given by the diagrams (here and in the following $p=(\omega,k)$)
\begin{equation}
\begin{split}
&G_{\varepsilon\varphi_a}^0(p) (-C(p))G_{\varphi_a\varepsilon}^0(p) =\\
&	  \begin{gathered}\includegraphics[width=0.1\linewidth,angle=0]{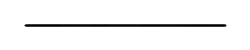}\end{gathered}
	+ \begin{gathered}\includegraphics[width=0.2\linewidth,angle=0]{fig/tadpole2}\end{gathered}
	+ \begin{gathered}\includegraphics[width=0.2\linewidth,angle=0]{fig/d1.png}\end{gathered}
	+ \left[\begin{gathered}\includegraphics[width=0.2\linewidth,angle=0]{fig/d4.png}\end{gathered}
	+ \hbox{c.c}\right]
\end{split}
\end{equation}
which gives 
\begin{equation}\label{eq_C_eval}
\begin{split}
C(\omega,k)
	&= 2\kappa T^2 k^2 
	+ 2cT^2\tilde \lambda' k^2 \int_{p'} G_{\varepsilon \varepsilon}(p')
	+ \frac12\lambda^2 k^4 \int_{p'} G_{\varepsilon \varepsilon}(p')G_{\varepsilon \varepsilon}(p+p')
	\\
	&+ \left[ 2i cT^2\lambda \tilde \lambda k^2\int_{p'} k\cdot k' G_{\varphi_a\varepsilon}(p')G_{\varepsilon \varepsilon}(p+p') + \hbox{c.c.}\right]\, .
\end{split}
\end{equation}
Loops with all $\omega'$ poles in the same half of the complex plane vanish. This is why terms such as e.g.~$\int_{p'} G_{\varphi_a \varepsilon}(p')G_{\varphi_a \varepsilon}(p+p') = 0$ do not appear in the above expression.

The diagrams contributing to the self-energy $\Sigma(p)$ are
\begin{equation}\label{eq_Sigma_def}
G_{\varepsilon \varphi_a}^0(p)\Sigma(p)G_{\varepsilon \varepsilon}^0(p)
	\equiv 
	\begin{gathered}\includegraphics[width=0.2\linewidth,angle=0]{fig/tadpole1}\end{gathered}
	+ \begin{gathered}\includegraphics[width=0.2\linewidth,angle=0]{fig/d2.png}\end{gathered}
	+ \begin{gathered}\includegraphics[width=0.2\linewidth,angle=0]{fig/d3.png}\end{gathered}
	\, ,
\end{equation}
which gives
\begin{equation}\label{eq_Sig_eval}
\begin{split}
\Sigma(p)
	&= -i\lambda' k^2 \int_{p'}G_{\varepsilon \varepsilon}(p')-\lambda^2 k^2 \int_{p'} k'^2 G_{\varphi_a\varepsilon}(p') G_{\varepsilon \varepsilon}(p+p')
	\\
	&+ icT^2 \lambda\tilde \lambda k^2 \int_{p'}k'\cdot(k+k') G_{\varphi_a \varepsilon}(p')G_{\varepsilon\varphi_a}(p+p')  \, .
\end{split}
\end{equation}
These diagrams get resummed, so that the full one-loop symmetric Green's function is 
\begin{equation}
G_{\varepsilon \varepsilon}(\omega,k)
	= \frac{C(\omega,k)}{\omega^2 + D^2 k^4 + 2D k^2 \Im \Sigma(\omega,k) + 2\omega \Re \Sigma(\omega,k)}\,. \label{eq:sym}
\end{equation}
One can recover the retarded Green's function from the symmetric Green's function (\ref{eq:sym}) using Kramers-Kronig relations. Parametrizing the retarded Green's function as
\begin{equation}
G^R_{\varepsilon \varepsilon}(\omega,k)
	= \frac{i \left(\kappa + \delta\kappa(\omega,k)\right) Tk^2}{\omega+ i D k^2 + \Sigma(\omega,k)}\, ,
\end{equation}
$\delta \kappa$ is related to $C$ above as
\begin{equation}
C = 2T^2 \kappa k^2 \left[1+ \frac{\Re \delta \kappa}{\kappa} + \frac{D k^2}{\omega} \frac{\Im \delta \kappa}{\kappa} + \frac{\Re \Sigma}{\omega}\right]\, .
\end{equation}
We will see that $\delta \kappa$ has a significantly simpler form than $C$.

Computing the integrals in \eqref{eq_Sigma_def} with a sharp momentum cutoff%
	\footnote{At this order in derivatives, all integrals over frequencies $\omega'$ are UV finite. Including higher order terms in derivatives however can lead to divergences that must be regulated consistently with the analytic structure of retarded and advanced correlators \cite{Gao:2018bxz}.} 
$\Lambda = 2\pi / \ell_{\rm th}$ one finds that the self-energy contains an analytic piece which is a correction to the diffusion constant and a non-analytic piece
\begin{equation}
\Sigma = i \delta D k^2 + \Sigma_\star\, .
\end{equation}
The correction to the diffusion constant is
\begin{equation}
\frac{\delta D}{D} = \frac{f_d}{c\, \ell_{\rm th}^d} \lambda_D\, ,
\end{equation}
with $f_d = {\rm Vol} (B_{d})  =  2,\, {\pi},\, \frac{4\pi}{3}$ for $d=1,2,3$ and where $\lambda_D$ is an effective dimensionless coupling given by
\begin{equation}
\lambda_D = - \frac{c^2 T^2}{2D^2} \left[\lambda(\lambda+\tilde \lambda) + 2\lambda' D\right]\, .
\end{equation}
Similar analytic contributions that are subleading in the cutoff have been dropped. The non-analytic part is cutoff independent and given by
\begin{equation}
\Sigma_\star(\omega,k)
	= \alpha_d(\omega,k) \frac{cT^2}{D^2} k^2 \left[ \omega \lambda(\lambda+\tilde \lambda) + iDk^2 \lambda\tilde \lambda\right]
	\equiv \alpha_d(\omega,k) k^2 f_\Sigma(\omega,k)\, , 
\end{equation}
where we factored out the part that depends on dimensionality
\begin{subequations}\label{eq_alpha_app}
\begin{align}
&&\alpha_1(\omega,k) &= \frac{1}{4} \left[k^2 - \frac{2i\omega}{D}\right]^{-1/2}\, , 
& (d=1)\\
&&\alpha_2(\omega,k) &= -\frac{1}{16\pi}\log \left[k^2 - \frac{2i\omega}{D}\right]\, , 
& (d=2)\\
&&\alpha_3(\omega,k) &= -\frac{1}{32\pi} \left[k^2 - \frac{2i\omega}{D}\right]^{1/2}\, .
& (d=3)
\end{align}
\end{subequations}
The correction to the numerator $\delta \kappa(\omega,k)$ can similarly be separated into an analytic part that corrects the thermal conductivity and a non-analytic part
\begin{equation}\label{eq_kappa_w}
\frac{\delta\kappa(\omega,k)}{\kappa}
	= \frac{f_d}{c\,\ell_{\rm th}^d}\lambda_\kappa
	 + \frac{1}{\kappa}\alpha_d(\omega,k) f_\kappa(k)
	 \, , \quad
	 \hbox{with} \quad 
	\lambda_\kappa = \frac{c^2 T^2 \tilde\lambda'}{D}\, , \quad
	f_\kappa(k) = \frac{cT^2}{D^2}k^2\lambda\tilde \lambda\, , 
\end{equation}
where $\alpha_d(\omega,k)$ is still given by \eqref{eq_alpha_app} and with $f_d = {\rm Vol} (B_{d})  =  2,\, {\pi},\, \frac{4\pi}{3}$ for $d=1,2,3$.

\section{Two diffusive charges}\label{app_2n}

This formalism can be straightforwardly generalized to account for more conservation laws. In this section we outline new features that arise in the case of multiple conserved scalar densities $n^I$, $I=1,2,\ldots , N$. The quadratic and cubic action are
\begin{subequations}
\begin{align}
\mathcal L_2
	&= iT \sigma_{IJ} \nabla\varphi_a^I \nabla \varphi_a^J - \varphi^I \left(\dot n_J - D_{JK} \nabla^2 n^K\right)\, , \\
\mathcal L_3
	&= -\lambda_{IJK} \nabla \varphi_a^I n^J \nabla n^K + \widetilde\lambda_{IJK} \nabla \varphi^I_a \nabla \varphi^J_a n^K  \, . 
\end{align}
\end{subequations}
The qualitative difference with the single charge case is that now $\lambda_{IJK}$ can contain a piece that is anti-symmetric under $J \leftrightarrow K$, so that this term can no longer be expressed as $\sim\nabla^2 \varphi_a n^2$ by integrating by parts \cite{PhysRevB.73.035113,Kovtun:2014nsa}. As a consequence, some of the factors of external momentum $k$ in \eqref{eq_C_eval} are replaced by the loop momentum $k'$, and the final expression is less $k^2$ suppressed. Specifically, this happens with the third diagram in \eqref{eq_C_eval}. One can easily check that no such enhancement occurs in the diagrams contributing to the self-energy \eqref{eq_Sig_eval}. As a result, only the conductivity has an enhanced $k^2\to0$ contribution compared to the single charge case \eqref{eq_kappa_w}. It is  proportional to $(\lambda_{I[JK]})^2$ and has the form
\begin{equation}\label{eq_cond_2n}
\delta \sigma(\omega,k)
	\sim {\omega} \, \alpha_d^{\rm mix}(\omega,k)
	+ \cdots \, ,
\end{equation}
where $\cdots$ denote contributions that are further $k^2$ suppressed (as in the single charge case), and 
\begin{subequations}\label{eq_alpha_mix}
\begin{align}
&&\alpha_1^{\rm mix}(\omega,k) &= \left[k^2 - \frac{2i\omega}{D_{\rm mix}}\right]^{-1/2}\, , 
& (d=1)\\
&&\alpha_2^{\rm mix}(\omega,k) &= \log \left[k^2 - \frac{2i\omega}{D_{\rm mix}}\right]\, , 
& (d=2)\\
&&\alpha_3^{\rm mix}(\omega,k) &= \left[k^2 - \frac{2i\omega}{D_{\rm mix}}\right]^{1/2}\, .
& (d=3)
\end{align}
\end{subequations}

\begin{figure}[h]
\centerline{
\subfigure{
\begin{overpic}[width=0.40\textwidth,tics=10]{fig/cut}
	 \put (55,47) {$\omega'=i{\color{inkblue}D_1} k'^2$}
	 \put (55,16) {$\omega+\omega'=-i{\color{inkred}D_2} (k+k')^2$}
	 \put (13,36) {$\omega,k$} 
	 \put (29,49) {$\omega',k'$} 
\end{overpic} 
}
\hspace{50pt}
\subfigure{
\begin{overpic}[width=0.35\textwidth,tics=10]{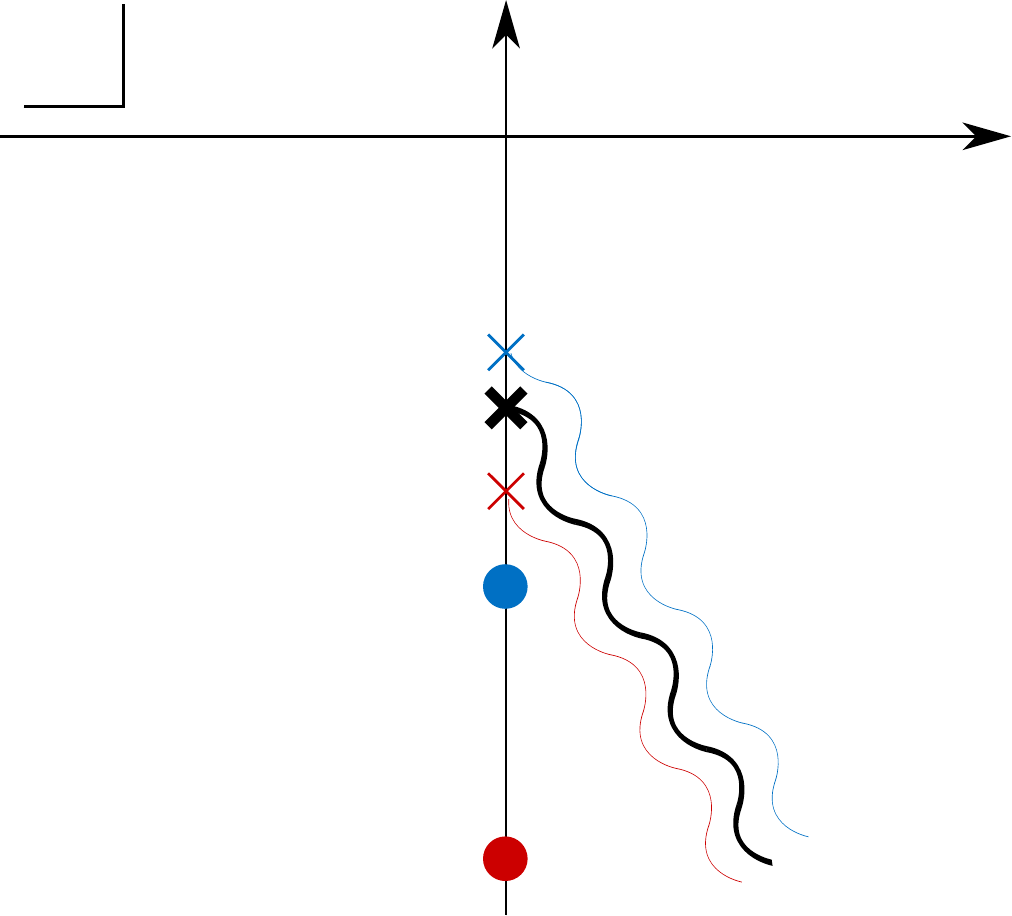}
	 \put (5,83) {$\omega$}
	 \put (10,48) {$ {- \frac{i}{2}D_{\rm mix} k^2}$} 
	 \put (17,30) {$ {\color{inkblue}- iD_1 k^2}$} 
	 \put (17,4) {$ {\color{inkred}- iD_2 k^2}$} 
\end{overpic}
}}
\caption{On-shell condition when two different diffusive poles are picked up in the loop (left), and analytic structure of $G^R_{nn}(\omega,k)$ in the presence of two coupled diffusive charges. The branch cuts have been rotated for clarity, and the splitting of the diffusive poles (see Fig.~\ref{fig_non_ana}) is not shown.
\label{fig_non_ana_2}}
\end{figure}
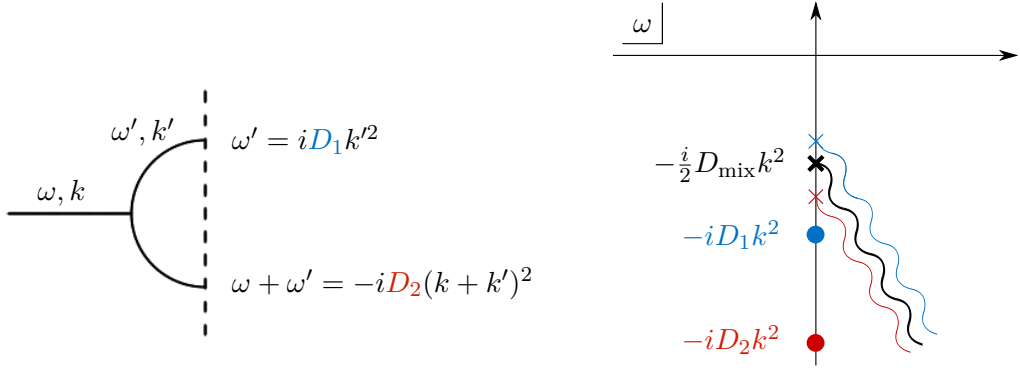

The branch point in this contribution occurs at $\omega_\star = -\frac{i}{2}D_{\rm mix} k^2$, with $D_{\rm mix} = 2(D_1^{-1}+D_2^{-1})^{-1}$, which can be understood as follows: the anti-symmetry of $\lambda_{I[JK]}$ requires two different propagators in the loop. Putting both legs on shell, as in Fig.~\ref{fig_non_ana_2}, gives
\begin{equation}
\omega_\star
	= -i \min_{k'} \left[ D_1 k'^2 + D_2(k^2 + 2k\cdot k' + k'^2)\right]
	= -i \frac{D_1D_2}{D_1+D_2} k^2\, .
\end{equation}
Here we have specialized to the case of $N=2$ interacting diffusive charges, and $D_1,\, D_2$ are the eigenvalues of $D_{IJ}$. The Green's function also contains branch points at $\omega_\star = -\frac{i}{2}D_{1}k^2$ and $ -\frac{i}{2}D_{2}k^2$ in analogy with the single charge case, see Fig.~\ref{fig_non_ana_2}.\newpage

Taking $k\to 0$ in \eqref{eq_cond_2n}, one finds that in the presence of mixing between two diffusive charges, the optical conductivity receives a correction $\delta\sigma(\omega) \sim |\omega|^{d/2}$ for $d=1$ and $d=3$, and $\sigma (\omega) = \omega\log \omega$ for $d=2$.

\end{document}